\documentclass{birkjour_t2}
\usepackage{amsmath}
\usepackage{amssymb}
\newcommand*{\defeq}{\mathrel{\vcenter{\baselineskip0.5ex \lineskiplimit0pt
            \hbox{\scriptsize.}\hbox{\scriptsize.}}}%
    =}
\usepackage{amsfonts}
\DeclareMathAlphabet{\mathbfit}{\encodingdefault}{\rmdefault}{bx}{sl}
\DeclareMathAlphabet{\mathbfsf}{\encodingdefault}{\sfdefault}{bx}{n}
\begin{document}

%
%
%
%
%
%
%
%
%

\title[Laplace Stretch]{Laplace Stretch: Eulerian and Lagrangian Formulations}

\author[Freed \textit{et~al.}]{Alan D.\ Freed}
    
\address{%
    Department of Mechanical Engineering, 
    Texas A\&\mbox{M} University,
    College Station, TX 77843, 
    United States \\
    and \\ 
    Impact Physics Branch, 
    U.S.\ Army Research Laboratory,
    Aberdeen Proving Ground, 
    Aberdeen, MD 21005, 
    United States}
    
\email{afreed@tamu.edu}

\author{Shahla Zamani}

\address{%
    Department of Mechanical Engineering, 
    Texas A\&\mbox{M} University,
    College Station, TX 77843, 
    United States}

\author{L{\'a}szl{\'o} Szab{\'o}}

\address{
    Budapest University of Technology and Economics,
    Budapest, H-1521, Hungary}

\author{John D.\ Clayton}

\address{%
    Impact Physics Branch, 
    U.S.\ Army Research Laboratory,
    Aberdeen Proving Ground, 
    Aberdeen, MD 21005, 
    United States}

    
    
\date{\today}

    
\begin{abstract}

Two triangular factorizations of the deformation gradient tensor are studied.  The first, termed the Lagrangian formulation, consists of an upper-triangular stretch premultiplied by a rotation tensor.  The second, termed the Eulerian formulation, consists of a lower-triangular stretch postmultiplied by a different rotation tensor.  The corresponding stretch tensors are denoted as the Lagrangian and Eulerian Laplace stretches, respectively.  Kinematics (with physical interpretations) and work conjugate stress measures are analyzed and compared for each formulation.    While the Lagrangian formulation has been used in prior work for constitutive modeling of anisotropic and hyper\-elastic materials, the Eulerian formulation, which may be advantageous for modeling isotropic solids and fluids with no physically identifiable reference configuration, does not seem to have been used elsewhere in a continuum mechanical setting.  
    
\end{abstract}
    
\maketitle
    

\section{Introduction}

The deformation gradient admits a number of different triangular decompositions, whereby in each case the full deformation gradient matrix is decomposed into a product of an orthogonal tensor and a triangular stretch tensor.  Restricting analysis to a
deformation gradient with positive determinant, each orthogonal tensor is a rotation, and each corresponding stretch, either upper or lower triangular, is unique for its corresponding rotation.  The first such decomposition considered herein splits the deformation gradient into an upper-triangular stretch followed (i.e., premultiplied) by a rotation tensor.  This kinematic construction is referred to here as the Lagrangian formulation of the triangular decomposition, also known as a Gram-Schmidt factorization.   The second such decomposition studied in this paper splits the deformation gradient tensor into a rotation tensor followed (premultiplied) by a lower-triangular stretch tensor.  This construction is referred to as the Eulerian formulation of the triangular decomposition of deformation.  

The Lagrangian triangular decomposition was first introduced in the context of continuum mechanics by McClellan \cite{McLellan76,McLellan80}.  The corresponding upper-triangular stretch tensor was proven very appealing for
modeling anisotropic hyperelastic materials by Srinivasa \cite{Srinivasa12}.  Other recent applications of the Lagrangian
decomposition address shape memory polymers \cite{GhoshSrinivasa14}, anisotropic composites \cite{ErelFreed17}, biological membranes \cite{Freedetal17}, and soft biological tissues \cite{ClaytonFreed20}.  Advantages and drawbacks of using the upper-triangular decomposition in constitutive models are discussed in these and related works \cite{FreedZamani19}.  Notably, the triangular decomposition, unlike the polar decomposition, requires no eigenvector analysis to invoke, and the components of stretch have an obvious physical interpretation that facilitates direct and unambiguous parameterization of constitutive response data.

In general, Lagrangian formulations (e.g., constitutive models based on Lagrangian measures of strain) are preferred for modeling anisotropic solids, as-well-as certain isotropic solids, that have a clearly defined initial, stress-free, or `reference' state.  This is readily apparent for single crystals, for example, whereby a reference state is identified with the regular lattice geometry occupied by atoms in their minimum energy (ground) state.  Hyper\-elasticity is usually invoked in this context \cite{McLellan80,Clayton11}, whereby an energy potential depending on a Lagrangian stretch tensor is prescribed.  Eulerian formulations, in contrast, are often preferred for modeling isotropic solids (and fluids) that have no obvious initial or reference state.  For example, many biological tissues, in vivo, are perpetually under tension, and a stress-free reference state is never physically realized.  Eulerian forms are also used for hypo\-elastic constitutive modeling that is often more popular than hyper\-elasticity for solving initial-boundary value problems numerically.  However, prior to the present work, no application of the Eulerian lower-triangular decomposition in the context of continuum mechanics seems to have been reported.  A different triangular decomposition of the deformation gradient was invoked by Souchet \cite{Souchet93}, consisting of a lower-triangular stretch premultiplied (rather than postmultiplied) by a rotation.  In that case, the lower-triangular stretch is considered a Lagrangian stretch measure rather than an Eulerian stretch measure, as newly studied herein.

\section{Deformation}

Consider a body $\mathcal{B}$ embedded in a three-dimensional, Euclidean, point space oriented against a triad of orthogonal, unit, base vectors $( \vec{\boldsymbol{\imath}}, \vec{\boldsymbol{\jmath}}, \vec{\boldsymbol{k}})$.  The motion $\boldsymbol{x} = \boldsymbol{\chi}(\boldsymbol{X},t)$ of some particle $\mathcal{P}$ located in $\mathcal{B}$ describes a homeomorphism that takes its original location $\boldsymbol{X} = X_1 \vec{\boldsymbol{\imath}} + X_2 \vec{\boldsymbol{\jmath}} + X_3 \vec{\boldsymbol{k}}$ belonging to the body's reference configuration $\kappa_r$ and places it into another location $\boldsymbol{x} = x_1 \vec{\boldsymbol{\imath}} + x_2 \vec{\boldsymbol{\jmath}} + x_3 \vec{\boldsymbol{k}}$ where $\mathcal{P}$ resides in the body's current configuration $\kappa_t$.  

For convenience, we write these two position vectors as $\boldsymbol{X} = X_i \, \vec{\mathbfsf{e}}_i$ and $\boldsymbol{x} = x_i \, \vec{\mathbfsf{e}}_i$ by selecting an indexing strategy, e.g., $( \vec{\boldsymbol{\imath}}, \vec{\boldsymbol{\jmath}}, \vec{\boldsymbol{k}}) \mapsto ( \vec{\mathbfsf{e}}_1, \vec{\mathbfsf{e}}_2, \vec{\mathbfsf{e}}_3 )$, to ensure that the 1~material direction and the 12~material surface embed with the motion, as they are invariant under transformations of Laplace stretch \cite{McLellan80}.  How to select an appropriate indexing strategy is the topic of Ref.~\cite{Pauletal20}.  This selection technique has been applied to our example problems. 

A deformation gradient $\mathbf{F}$ maps the set of all tangent vectors located at particle $\mathcal{P}$ in body $\mathcal{B}$ from its reference configuration $\kappa_r$ into the current configuration $\kappa_t$. We assume that a body is simply connected and its motion $\boldsymbol{\chi}$ is sufficiently differentiable so that $\mathbf{F} = \partial \boldsymbol{\chi}(\boldsymbol{X},t) / \partial \boldsymbol{X}$ exists and therefore
\begin{equation}
\label{deformationVectors}
F_{ij} = \frac{\partial \chi_i (\boldsymbol{X},t)}{\partial X_j} =
\begin{bmatrix}
F_{11} & F_{12} & F_{13}\\
F_{21} & F_{22} & F_{23}\\
F_{31} & F_{32} & F_{33}
\end{bmatrix} = 
\left[ \begin{tabular}{c}
\ensuremath{\boldsymbol{f}^r_1} \\ \hline
\ensuremath{\boldsymbol{f}^r_2} \\ \hline
\ensuremath{\boldsymbol{f}^r_3} \end{tabular} \right] = 
\left[ \begin{array}{c|c|c}
\boldsymbol{f}^c_1 & \boldsymbol{f}^c_2 & \boldsymbol{f}^c_3
\end{array} \right]
\end{equation}
where vectors $\boldsymbol{f}^r_i = F_{ij} \, \vec{\mathbfsf{e}}_j$ contain the rows of tensor $\mathbf{F} = F_{ij} \, \vec{\mathbfsf{e}}_i \otimes \vec{\mathbfsf{e}}_j$, while vectors $\boldsymbol{f}^c_i = F_{ji} \, \vec{\mathbfsf{e}}_j$ contain its columns, $i=1,2,3$, with repeated indices being summed according to Einstein's summation convention.  

It follows straightaway that the right, Cauchy-Green, deformation tensor $\mathbf{C} \defeq \mathbf{F}^{\mathsf{T}} \mathbf{F} = C_{ij} \, \vec{\mathbfsf{e}}_i \otimes \vec{\mathbfsf{e}}_j$, which is a Lagrangian description of deformation, has components of
\begin{equation}
    C_{ij} = \begin{bmatrix}
    \boldsymbol{f}^c_1 \cdot \boldsymbol{f}^c_1 &
       \boldsymbol{f}^c_1 \cdot \boldsymbol{f}^c_2 &
       \boldsymbol{f}^c_1 \cdot \boldsymbol{f}^c_3 \\
    \boldsymbol{f}^c_2 \cdot \boldsymbol{f}^c_1 &
       \boldsymbol{f}^c_2 \cdot \boldsymbol{f}^c_2 &
       \boldsymbol{f}^c_2 \cdot \boldsymbol{f}^c_3 \\
    \boldsymbol{f}^c_3 \cdot \boldsymbol{f}^c_1 &
       \boldsymbol{f}^c_3 \cdot \boldsymbol{f}^c_2 &
       \boldsymbol{f}^c_3 \cdot \boldsymbol{f}^c_3 
    \end{bmatrix}
    \tag{1.2a}
    \label{rightCauchyGreen}
\end{equation}
while the left, Cauchy-Green, deformation tensor $\mathbf{B} \defeq \mathbf{FF}^{\mathsf{T}} = B_{ij} \, \vec{\mathbfsf{e}}_i \otimes \vec{\mathbfsf{e}}_j$, which is an Eulerian description of deformation, has components of
\begin{equation}
B_{ij} = \begin{bmatrix}
\boldsymbol{f}^r_1 \cdot \boldsymbol{f}^r_1 &
\boldsymbol{f}^r_1 \cdot \boldsymbol{f}^r_2 &
\boldsymbol{f}^r_1 \cdot \boldsymbol{f}^r_3 \\
\boldsymbol{f}^r_2 \cdot \boldsymbol{f}^r_1 &
\boldsymbol{f}^r_2 \cdot \boldsymbol{f}^r_2 &
\boldsymbol{f}^r_2 \cdot \boldsymbol{f}^r_3 \\
\boldsymbol{f}^r_3 \cdot \boldsymbol{f}^r_1 &
\boldsymbol{f}^r_3 \cdot \boldsymbol{f}^r_2 &
\boldsymbol{f}^r_3 \cdot \boldsymbol{f}^r_3 
\end{bmatrix} 
\tag{1.2b}
\label{leftCauchyGreen}
\end{equation}
both of which are symmetric because, for example, $\boldsymbol{f}^r_1 \cdot \boldsymbol{f}^r_2 = \boldsymbol{f}^r_2 \cdot \boldsymbol{f}^r_1$ where $\boldsymbol{f}^r_1 \cdot \boldsymbol{f}^r_2 = F_{1i} F_{2i} = F_{11} F_{21} + F_{12} F_{22} + F_{13} F_{23}$, etc.

\section{Laplace Stretch}

Laplace stretch, as it has been used in the literature to date, e.g., \cite{McLellan76,McLellan80,Srinivasa12,GhoshSrinivasa14,FreedSrinivasa15,ErelFreed17,Freedetal17,FreedZamani18,FreedZamani19,Freedetal19,Pauletal20,ClaytonFreed20}, derives from a Gram-Schmidt (or \textbf{QR}) decomposition of the deformation gradient $\mathbf{F}$, where matrix \textbf{Q} is orthogonal, and matrix \textbf{R} is upper triangular.  

Given a coordinate system with base vectors $( \vec{\mathbfsf{e}}_1 , \vec{\mathbfsf{e}}_2 , \vec{\mathbfsf{e}}_3 )$, we denote such a decomposition as $\mathbf{F} = \boldsymbol{\mathcal{RU}}$, where $\boldsymbol{\mathcal{R}} = \mathcal{R}_{ij} \, \vec{\mathbfsf{e}}_i \otimes \vec{\mathbfsf{e}}_j$ has orthogonal components, and $\boldsymbol{\mathcal{U}} = \mathcal{U}_{ij} \, \vec{\mathbfsf{e}}_i \otimes \vec{\mathbfsf{e}}_j$ has upper-triangular components.  We select this calligraphic notation to illustrate its similarities and differences with the common polar decomposition $\mathbf{F} = \mathbf{RU}$, where $\mathbf{R} = R_{ij} \, \vec{\mathbfsf{e}}_i \otimes \vec{\mathbfsf{e}}_j$ has orthogonal components, and $\mathbf{U} = U_{ij} \, \vec{\mathbfsf{e}}_i \otimes \vec{\mathbfsf{e}}_j$ has symmetric components.  Lagrangian fields $\mathbf{U}$ and $\boldsymbol{\mathcal{U}}$ are distinct measures for stretch.

A polar decomposition of the deformation gradient, $\mathbf{F} = \mathbf{RU} = \mathbf{VR}$, produces a Lagrangian measure for stretch (the right-stretch tensor $\mathbf{U}$) and an Eulerian measure for stretch (the left-stretch tensor $\mathbf{V}$) that share in a common, orthogonal, rotation tensor $\mathbf{R}$. An objective of this document is to develop an Eulerian measure for stretch whose components populate a triangular matrix such that $\mathbf{F} = \boldsymbol{\mathcal{R}}^{\!L} \boldsymbol{\mathcal{U}} = \boldsymbol{\mathcal{VR}}^{\!E}$, where $\boldsymbol{\mathcal{U}}$ is the Lagrangian Laplace stretch, and where $\boldsymbol{\mathcal{V}}$ is the Eulerian Laplace stretch, both with triangular elements.  In contrast with the polar rotation $\mathbf{R}$, the Lagrangian $\boldsymbol{\mathcal{R}}^{\!L}$ and Eulerian $\boldsymbol{\mathcal{R}}^{\!E}$ Gram rotations are distinct rotations.  The Laplace stretches therefore relate via $\boldsymbol{\mathcal{U}} = \boldsymbol{\mathcal{R}}^{\!L^{\mathsf{T}}} \boldsymbol{\mathcal{V}} \boldsymbol{\mathcal{R}}^{\!E}$ and $\boldsymbol{\mathcal{V}} = \boldsymbol{\mathcal{R}}^{\!L} \boldsymbol{\mathcal{U}} \boldsymbol{\mathcal{R}}^{\!E^{\mathsf{T}}}$.   

\subsection{Lagrangian Laplace Stretch}

Here we describe a Gram-Schmidt factorization of the deformation gradient, i.e., $\mathbf{F} = \boldsymbol{\mathcal{R}}^{\!L} \boldsymbol{\mathcal{U}}$, wherein $\boldsymbol{\mathcal{U}} = \mathcal{U}_{ij} \, \vec{\mathbfsf{e}}_i \otimes \vec{\mathbfsf{e}}_j$ is called the Lagrangian Laplace stretch, or the right Laplace stretch.

Srinivasa \cite{Srinivasa12} applied a Cholesky decomposition to the symmetric, positive-definite, right, Cauchy-Green, deformation tensor $\mathbf{C}$ to establish the components of his stretch tensor, denoted here as $\boldsymbol{\mathcal{U}} = \mathcal{U}_{ij} \,  \vec{\mathbfsf{e}}_i \otimes \vec{\mathbfsf{e}}_j$; in particular,\footnote{
   Regarding Lagrangian stretches with triangular elements, McLellan \cite{McLellan76,McLellan80} was the first to propose an upper-triangular decomposition of the deformation gradient.  Later, Souchet \cite{Souchet93} constructed a stretch tensor with lower-triangular components.  We use Srinivasa's \cite{Srinivasa12} approach for populating an upper-triangular stretch because, of these three Lagrangian approaches, his is the simplest framework to apply.
}
\begin{equation}
    \begin{aligned}
    \mathcal{U}_{11} & = \sqrt{C_{11}} & 
       \mathcal{U}_{12} & = C_{12} / \mathcal{U}_{11} &
       \mathcal{U}_{13} & = C_{13} / \mathcal{U}_{11} \\
    \mathcal{U}_{21} & = 0 &
       \mathcal{U}_{22} & = \sqrt{C_{22} - \mathcal{U}_{12}^{\,2}} &
       \mathcal{U}_{23} & = \bigl( C_{23} - \mathcal{U}_{12\,}\mathcal{U}_{13} \bigr) / \mathcal{U}_{22} \\
    \mathcal{U}_{31} & = 0 &
       \mathcal{U}_{32} & = 0 & 
       \mathcal{U}_{33} & = \sqrt{C_{33} - \mathcal{U}_{13}^{\,2} - \mathcal{U}_{23}^{\,2}}
    \end{aligned}
    \label{LagrangianLaplaceStretch}
\end{equation}
where components of the Lagrangian Laplace stretch $\mathcal{U}_{ij}$ are upper triangular.  Its inverse $\boldsymbol{\mathcal{U}}^{-1} = \mathcal{U}^{-1}_{ij} \, \vec{\mathbfsf{e}}_i \otimes \vec{\mathbfsf{e}}_j$ follows straightaway, having components of
\begin{equation}
    \mathcal{U}^{-1}_{ij} = \begin{bmatrix}
    1 / \mathcal{U}_{11} & -\mathcal{U}_{12} / \mathcal{U}_{11} \mathcal{U}_{22} & 
       ( \mathcal{U}_{12} \mathcal{U}_{23} - \mathcal{U}_{13} \mathcal{U}_{22} ) / 
       \mathcal{U}_{11} \mathcal{U}_{22} \mathcal{U}_{33} \\
    0 & 1 / \mathcal{U}_{22} & -\mathcal{U}_{23} / \mathcal{U}_{22} \mathcal{U}_{33} \\
    0 & 0 & 1 / \mathcal{U}_{33}
    \end{bmatrix} 
    \label{inverseLagrangianLaplaceStretch}
\end{equation}
thereby requiring that each $\mathcal{U}_{ii}$, no sum on $i$, to be positive---a condition satisfied because of mass conservation.  It is easily shown that the Lagrangian Laplace stretch $\mathcal{U}_{ij}$ belongs to a group under the operation of matrix multiplication.  This group is comprised of all real, $3 \! \times \! 3$, upper-triangular matrices with positive diagonal elements \cite{McLellan80}.  Having a stretch tensor with this property has proven to be useful in applications, e.g., \cite{McLellan80,Freedetal19}, as it does here.

A Gram factorization of the deformation gradient $\mathbf{F} = F_{ij} \, \vec{\mathbfsf{e}}_i \otimes \vec{\mathbfsf{e}}_j$ produces a Lagrangian rotation tensor $\boldsymbol{\mathcal{R}}^{\!L} = \delta_{ij} \, 
\vec{\mathbfsf{e}}^L_i \otimes \vec{\mathbfsf{e}}_j = \mathcal{R}^{\!L}_{ij} \, 
\vec{\mathbfsf{e}}_i \otimes \vec{\mathbfsf{e}}_j$ described by
\begin{subequations}
    \label{LagrangianRotation}
    \begin{align}
    \mathcal{R}^{\!L}_{ij} & = 
    \left[ \begin{array}{c|c|c}
    \vec{\mathbfsf{e}}^L_1 & \vec{\mathbfsf{e}}^L_2 & \vec{\mathbfsf{e}}^L_3
    \end{array} \right] \\
    \intertext{whose columns constitute unit base vectors that can be constructed via}
    \vec{\mathbfsf{e}}^L_1 & \defeq \frac{\boldsymbol{f}^c_1}{\| \boldsymbol{f}^c_1 \|} \\
    \vec{\mathbfsf{e}}^L_2 & \defeq \frac{\boldsymbol{f}^c_2 - 
       ( \boldsymbol{f}^c_2 \cdot \vec{\mathbfsf{e}}^L_1 ) \vec{\mathbfsf{e}}^L_1} 
       {\| \boldsymbol{f}^c_2 - 
       ( \boldsymbol{f}^c_2 \cdot \vec{\mathbfsf{e}}^L_1 ) \vec{\mathbfsf{e}}^L_1 \|} \\
    \vec{\mathbfsf{e}}^L_3 & \defeq \frac{\boldsymbol{f}^c_3 - 
       ( \boldsymbol{f}^c_3 \cdot \vec{\mathbfsf{e}}^L_1 ) \vec{\mathbfsf{e}}^L_1 - \
       ( \boldsymbol{f}^c_3 \cdot \vec{\mathbfsf{e}}^L_2 ) \vec{\mathbfsf{e}}^L_2 } {\| \boldsymbol{f}^c_3 - 
       ( \boldsymbol{f}^c_3 \cdot \vec{\mathbfsf{e}}^L_1 ) \vec{\mathbfsf{e}}^L_1 - \
       ( \boldsymbol{f}^c_3 \cdot \vec{\mathbfsf{e}}^L_2 ) \vec{\mathbfsf{e}}^L_2  \|}
    \end{align}
\end{subequations}
wherein Laplace's technique for removing successive orthogonal projections \cite{Leonetal13} is apparent, with norm $\| \boldsymbol{f}^c_1 \| \defeq \sqrt{\boldsymbol{f}^c_1 \cdot \boldsymbol{f}^c_1}$, etc.  It therefore follows that the Lagrangian Laplace stretch has components which can be expressed as
\begin{equation}
    \mathcal{U}_{ij} = \begin{bmatrix}
    \vec{\mathbfsf{e}}^L_1 \cdot \boldsymbol{f}^c_1 &
       \vec{\mathbfsf{e}}^L_1 \cdot \boldsymbol{f}^c_2 &
       \vec{\mathbfsf{e}}^L_1 \cdot \boldsymbol{f}^c_3 \\
    0 & \vec{\mathbfsf{e}}^L_2 \cdot \boldsymbol{f}^c_2 &
       \vec{\mathbfsf{e}}^L_2 \cdot \boldsymbol{f}^c_3 \\
    0 & 0 &
       \vec{\mathbfsf{e}}^L_3 \cdot \boldsymbol{f}^c_3 
    \end{bmatrix}
    \label{geometricLagrangianLaplaceStretch}
\end{equation}
that provide a means of geometric interpretation for this measure of stretch.  Components $\mathcal{U}_{ij}$ of the Lagrangian Laplace stretch $\boldsymbol{\mathcal{U}} = \mathcal{U}_{ij} \, \vec{\mathbfsf{e}}_i \otimes \vec{\mathbfsf{e}}_j$ evaluated in a reference frame $( \vec{\mathbfsf{e}}_1 , \vec{\mathbfsf{e}}_2 , \vec{\mathbfsf{e}}_3 )$ are also projections of column vectors $\boldsymbol{f}^c_i$ extracted from a deformation gradient $\mathbf{F} = F_{ij} \, \vec{\mathbfsf{e}}_i \otimes \vec{\mathbfsf{e}}_j$ that are projected onto its Lagrangian coordinate axes $( \vec{\mathbfsf{e}}^L_1 , \vec{\mathbfsf{e}}^L_2 , \vec{\mathbfsf{e}}^L_3 )$.

\subsection{Eulerian Laplace Stretch}

Now we describe a Gram-Schmidt like factorization of the deformation gradient, viz., $\mathbf{F} = \boldsymbol{\mathcal{VR}}^{\!E}$, wherein $\boldsymbol{\mathcal{V}} = \mathcal{V}_{ij} \, \vec{\mathbfsf{e}}_i \otimes \vec{\mathbfsf{e}}_j$ is called the Eulerian Laplace stretch, or the left Laplace stretch.

Applying a Cholesky factorization to the symmetric, positive-definite, left, Cauchy-Green, deformation tensor $\mathbf{B} \defeq \mathbf{FF}^{\mathsf{T}} = \boldsymbol{\mathcal{VV}}^{\mathsf{T}}$ with components $\mathbf{B} = B_{ij} \, \vec{\mathbfsf{e}}_i \otimes \vec{\mathbfsf{e}}_j$ one can construct a stretch tensor $\boldsymbol{\mathcal{V}} = \mathcal{V}_{ij} \, \vec{\mathbfsf{e}}_i \otimes \vec{\mathbfsf{e}}_j$ whereby
\begin{equation}
    \begin{aligned}
    \mathcal{V}_{11} & = \sqrt{B_{11}} & \mathcal{V}_{12} & = 0 & 
       \mathcal{V}_{13} & = 0 \\
    \mathcal{V}_{21} & = B_{21} / \mathcal{V}_{11} & 
       \mathcal{V}_{22} & = \sqrt{B_{22} - \mathcal{V}_{21}^{\,2}} & 
       \mathcal{V}_{23} & = 0 \\
    \mathcal{V}_{31} & = B_{31} / \mathcal{V}_{11} &
       \mathcal{V}_{32} & = \bigl( B_{32} - \mathcal{V}_{21} 
       \mathcal{V}_{31} \bigr) / \mathcal{V}_{22} & 
       \mathcal{V}_{33} & = \sqrt{B_{33} - \mathcal{V}_{31}^{\,2} -
       \mathcal{V}_{32}^{\,2} }
    \end{aligned}
\label{EulerianLaplaceStretch}
\end{equation}
where we now select the lower-triangular matrix from the Cholesky decomposition to quantify the components of our new stretch tensor.  Its inverse $\boldsymbol{\mathcal{V}}^{-1} = \mathcal{V}^{-1}_{ij} \, \vec{\mathbfsf{e}}_i \otimes \vec{\mathbfsf{e}}_j$ follows straightaway, it having components of
\begin{equation}
\mathcal{V}^{-1}_{ij} = \begin{bmatrix}
1 / \mathcal{V}_{11} & 0 & 0 \\
-\mathcal{V}_{21} / \mathcal{V}_{11} \mathcal{V}_{22} & 1 / \mathcal{V}_{22} & 0 \\
( \mathcal{V}_{32} \mathcal{V}_{21} - \mathcal{V}_{31} \mathcal{V}_{22} ) / 
\mathcal{V}_{11} \mathcal{V}_{22} \mathcal{V}_{33} & -\mathcal{V}_{32} / \mathcal{V}_{22} \mathcal{V}_{33} & 1 / \mathcal{V}_{33}
\end{bmatrix} 
\label{inverseEulerianLaplaceStretch}
\end{equation}
thereby requiring each $\mathcal{V}_{ii}$, no sum on $i$, to be positive---a condition satisfied because of mass conservation.  It is easily shown that the Eulerian Laplace stretch $\mathcal{V}_{ij}$ belongs to a group under the operation of matrix multiplication. This group is comprised of all real, $3 \! \times \! 3$, lower-triangular matrices with positive diagonal elements.  The Eulerian and Lagrangian Laplace stretches belong to different mathematical groups.

A Gram-like\footnote{
   The Gram factorization of a square matrix results in an orthogonal matrix and an upper-triangular matrix.  Here we apply the same strategy, but we secure a different orthogonal matrix and a lower-triangular matrix; hence, the terminology `Gram like'.
} 
factorization of the deformation gradient $\mathbf{F} = F_{ij} \, \vec{\mathbfsf{e}}_i \otimes \vec{\mathbfsf{e}}_j$ can also describe an Eulerian rotation tensor $\boldsymbol{\mathcal{R}}^{\!E} = \delta_{ij} \, 
\vec{\mathbfsf{e}}_i \otimes \vec{\mathbfsf{e}}^E_j = \mathcal{R}^{\!E}_{ij} \, 
\vec{\mathbfsf{e}}_i \otimes \vec{\mathbfsf{e}}_j$ constructed as
\begin{subequations}
    \label{EulerianRotation}
    \begin{align}
    \mathcal{R}^{\!E}_{ij} & = 
    \left[ \begin{tabular}{c}
    \ensuremath{\vec{\mathbfsf{e}}^E_1} \\ \hline
    \ensuremath{\vec{\mathbfsf{e}}^{E^{\vphantom{|}}}_2} \\ \hline
    \ensuremath{\vec{\mathbfsf{e}}^{E^{\vphantom{|}}}_3} \end{tabular} \right] =
    \left[ \begin{array}{c|c|c}
    \vec{\mathbfsf{e}}^E_1 & \vec{\mathbfsf{e}}^E_2 & \vec{\mathbfsf{e}}^E_3
    \end{array} \right]^{\mathsf{T}} \\
    \intertext{whose rows constitute unit base vectors that can be constructed via}
    \vec{\mathbfsf{e}}^E_1 & \defeq \frac{\boldsymbol{f}^r_1}{\| \boldsymbol{f}^r_1 \|} \\
    \vec{\mathbfsf{e}}^E_2 & \defeq \frac{\boldsymbol{f}^r_2 - 
        ( \boldsymbol{f}^r_2 \cdot \vec{\mathbfsf{e}}^E_1 ) \vec{\mathbfsf{e}}^E_1} 
    {\| \boldsymbol{f}^r_2 - 
        ( \boldsymbol{f}^r_2 \cdot \vec{\mathbfsf{e}}^E_1 ) \vec{\mathbfsf{e}}^E_1 \|} \\
    \vec{\mathbfsf{e}}^E_3 & \defeq \frac{\boldsymbol{f}^r_3 - 
        ( \boldsymbol{f}^r_3 \cdot \vec{\mathbfsf{e}}^E_1 ) \vec{\mathbfsf{e}}^E_1 - \
        ( \boldsymbol{f}^r_3 \cdot \vec{\mathbfsf{e}}^E_2 ) \vec{\mathbfsf{e}}^E_2 } {\| \boldsymbol{f}^r_3 - 
        ( \boldsymbol{f}^r_3 \cdot \vec{\mathbfsf{e}}^E_1 ) \vec{\mathbfsf{e}}^E_1 - \
        ( \boldsymbol{f}^r_3 \cdot \vec{\mathbfsf{e}}^E_2 ) \vec{\mathbfsf{e}}^E_2  \|}
    \end{align}
\end{subequations}
where, again, Laplace's solution strategy for removing successive orthogonal projections \cite{Leonetal13} is apparent.  It follows that the Eulerian Laplace stretch has components which can be expressed as
\begin{equation}
\mathcal{V}_{ij} = \begin{bmatrix}
\boldsymbol{f}^r_1 \cdot \vec{\mathbfsf{e}}^E_1 & 0 & 0 \\
\boldsymbol{f}^r_2 \cdot \vec{\mathbfsf{e}}^E_1 &
\boldsymbol{f}^r_2 \cdot \vec{\mathbfsf{e}}^E_2 & 0 \\
\boldsymbol{f}^r_3 \cdot \vec{\mathbfsf{e}}^E_1 & 
\boldsymbol{f}^r_3 \cdot \vec{\mathbfsf{e}}^E_2 &
\boldsymbol{f}^r_3 \cdot \vec{\mathbfsf{e}}^E_3 
\end{bmatrix}
\label{geometricEulerianLaplaceStretch}
\end{equation}
that provide a means of geometric interpretation for this measure of stretch. Components $\mathcal{V}_{ij}$ of the Eulerian Laplace stretch $\boldsymbol{\mathcal{V}} = \mathcal{V}_{ij} \, \vec{\mathbfsf{e}}_i \otimes \vec{\mathbfsf{e}}_j$ evaluated in a reference frame $( \vec{\mathbfsf{e}}_1 , \vec{\mathbfsf{e}}_2 , \vec{\mathbfsf{e}}_3 )$ are also projections of row vectors $\boldsymbol{f}^r_i$ extracted from a deformation gradient $\mathbf{F} = F_{ij} \, \vec{\mathbfsf{e}}_i \otimes \vec{\mathbfsf{e}}_j$ that are projected onto its Eulerian coordinate axes $( \vec{\mathbfsf{e}}^E_1 , \vec{\mathbfsf{e}}^E_2 , \vec{\mathbfsf{e}}^E_3 )$.

Obviously, rotations $\boldsymbol{\mathcal{R}}^{\!L}$ and  $\boldsymbol{\mathcal{R}}^{\!E}$ are distinct, as are stretches $\boldsymbol{\mathcal{U}}$ and $\boldsymbol{\mathcal{V}}$, given that the deformation gradient $\mathbf{F}$ decomposes as $\mathbf{F} = \boldsymbol{\mathcal{R}}^{\!L} \boldsymbol{\mathcal{U}} = \boldsymbol{\mathcal{VR}}^{\!E}$ whose stretch tensors have triangular components $\mathcal{U}_{ij}$ and $\mathcal{V}_{ij}$.

\section{Physical Interpretation of Laplace Stretch Components}

Each Laplace stretch has six, independent, physical attributes.  There are three, ortho\-gonal, elongation ratios $a$, $b$ and $c$, and there are three, ortho\-gonal, simple shears $\alpha$, $\beta$ and $\gamma$.  Their Lagrangian interpretations are quantified in a coordinate system with base vectors $( \vec{\mathbfsf{e}}^L_1 , \vec{\mathbfsf{e}}^L_2 , \vec{\mathbfsf{e}}^L_3 )$, and are distinguished with an underline, viz., $\underline{a}$, $\underline{b}$, $\underline{c}$, $\underline{\alpha}$, $\underline{\beta}$ and $\underline{\gamma}$.  Their Eulerian interpretations are quantified in a coordinate system with base vectors $( \vec{\mathbfsf{e}}^E_1 , \vec{\mathbfsf{e}}^E_2 , \vec{\mathbfsf{e}}^E_3 )$, and are distinguished with an overline, viz., $\overline{a}$, $\overline{b}$, $\overline{c}$, $\overline{\alpha}$, $\overline{\beta}$ and $\overline{\gamma}$. In general, Lagrangian stretch attributes are distinct from their Eulerian counterparts.  However, their geometric interpretations are the same.  They differ only in their coordinate systems through which they are evaluated.

\subsection{Lagrangian Stretch Attributes}

The Lagrangian Laplace stretch has geometric interpretations that arise from Eqn.~(\ref{geometricLagrangianLaplaceStretch}) whereby one can assign \cite{FreedSrinivasa15}
\begin{subequations}
    \label{LagrangianPhysicalStretch}
    \begin{align}
\mathcal{U}_{ij} & = \begin{bmatrix}
\underline{a} & \underline{a} \underline{\gamma} & \underline{a} \underline{\beta} \\
0 & \underline{b} & \underline{b} \underline{\alpha} \\
0 & 0 & \underline{c} \end{bmatrix} = \begin{bmatrix}
\underline{a} & 0 & 0 \\
0 & \underline{b} & 0 \\
0 & 0 & \underline{c}
\end{bmatrix} \begin{bmatrix}
1 & 0 & \underline{\beta} \\
0 & 1 & \underline{\alpha} \\
0 & 0 & 1
\end{bmatrix} \begin{bmatrix}
1 & \underline{\gamma} & 0 \\
0 & 1 & 0 \\
0 & 0 & 1
\end{bmatrix} \\
\intertext{with an inverse of}
\mathcal{U}^{-1}_{ij} & = \begin{bmatrix}
1/\underline{a} & -\underline{\gamma} / \underline{b} & 
- ( \underline{\beta} - \underline{\alpha} \underline{\gamma} ) / \underline{c} \\
0 & 1/\underline{b} & -\underline{\alpha} / \underline{c} \\
0 & 0 & 1/\underline{c} \end{bmatrix}
\end{align}
\end{subequations}
whose constituents are measured in a coordinate frame with base vectors \cite{FreedZamani18}
\begin{subequations}
    \label{LagrangianBaseVectors}
    \begin{align}
    \vec{\mathbfsf{e}}^L_1 & = \boldsymbol{f}^c_1 \bigm/ \underline{a} \\
    \vec{\mathbfsf{e}}^L_2 & = \bigl( \boldsymbol{f}^c_2 - 
    \underline{\gamma} \boldsymbol{f}^c_1 \bigr) \bigm/ \underline{b} \\
    \vec{\mathbfsf{e}}^L_3 & = \bigl( \boldsymbol{f}^c_3 - 
    \underline{\alpha} \boldsymbol{f}^c_2 - 
    ( \underline{\beta} - \underline{\alpha} \underline{\gamma} ) \boldsymbol{f}^c_1 \bigr) \bigm/ \underline{c} 
    \end{align}
\end{subequations}
all of which are described in terms of physical attributes defined as
\begin{equation}
\underline{a} \defeq \mathcal{U}_{11} , \quad
\underline{b} \defeq \mathcal{U}_{22} , \quad
\underline{c} \defeq \mathcal{U}_{33} , \quad
\underline{\alpha} \defeq \frac{\mathcal{U}_{23}}{\mathcal{U}_{22}} , \quad
\underline{\beta} \defeq \frac{\mathcal{U}_{13}}{\mathcal{U}_{11}} , \quad
\underline{\gamma} \defeq \frac{\mathcal{U}_{12}}{\mathcal{U}_{11}}
\label{LagrangianPhysicalAttributes}
\end{equation}
where $\underline{a}$, $\underline{b}$ and $\underline{c}$ are elongations, while $\underline{\alpha}$, $\underline{\beta}$ and $\underline{\gamma}$ are magnitudes of shear, i.e., they are the extents of shear at unit elongation.  From conservation of mass, the elongations must be positive ($\underline{a} \in \mathbb{R}_+$, $\underline{b} \in \mathbb{R}_+$, $\underline{c} \in \mathbb{R}_+$), while the shears may be of either sign ($\underline{\alpha} \in \mathbb{R}$, $\underline{\beta} \in \mathbb{R}$, $\underline{\gamma} \in \mathbb{R}$).

\begin{figure}
    \centering
    \includegraphics[width=0.9\linewidth]{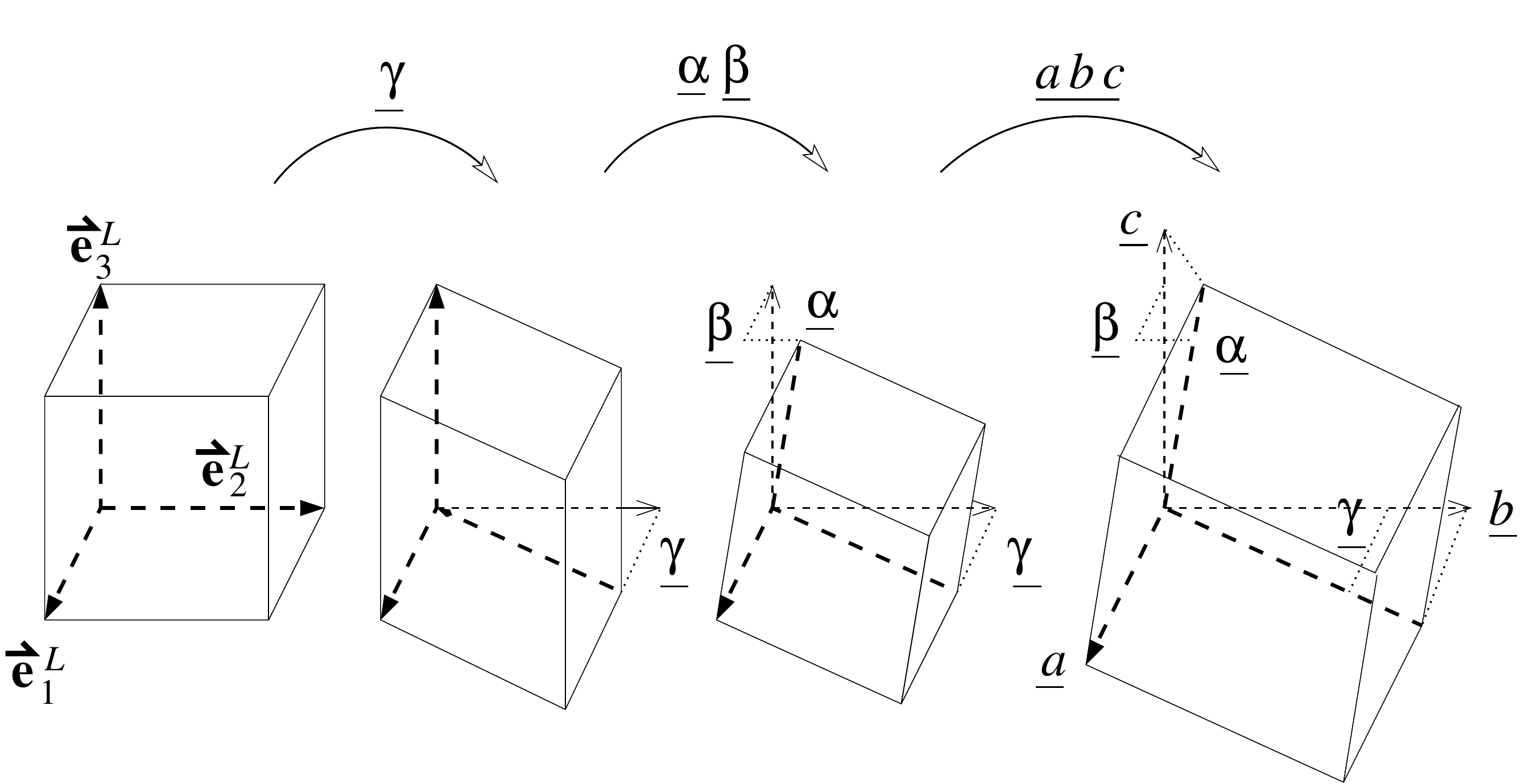}
    \label{fig:LagrangianStretch}
    \caption{A geometric interpretation for Lagrangian Laplace stretch.}
\end{figure}

According to Eqn.~(\ref{LagrangianPhysicalStretch}), the Lagrangian Laplace stretch arises from the following sequence of deformations: it starts with an in-plane shear $\underline{\gamma}$, followed by two out-of-plane shears $\underline{\alpha}$ and $\underline{\beta}$, and then finishes with three elongations $\underline{a}$, $\underline{b}$ and $\underline{c}$, as illustrated in Fig.~\ref{fig:LagrangianStretch}. Two vectors remain invariant under mappings of the Lagrangian Laplace stretch; they are: vector $\vec{\mathbfsf{e}}^L_1$ establishes the direction of in-plane shear, while vector $\vec{\mathbfsf{e}}^L_1 \times \vec{\mathbfsf{e}}^L_2$ points normal to the plane of in-plane shear \cite{McLellan80}.

\subsection{Eulerian Stretch Attributes}

The Eulerian Laplace stretch has geometric interpretations that arise from Eqn.~(\ref{geometricEulerianLaplaceStretch}) whereby one can assign
\begin{subequations}
    \label{EulerianPhysicalStretch}
    \begin{align}
\mathcal{V}_{ij} & = \begin{bmatrix}
\overline{a} & 0 & 0 \\
\overline{a} \overline{\gamma} & \overline{b} & 0 \\
\overline{a} \overline{\beta} & \overline{b} \overline{\alpha} & \overline{c} \end{bmatrix} = \begin{bmatrix}
1 & 0 & 0 \\
\overline{\gamma} & 1 & 0 \\
0 & 0 & 1
\end{bmatrix} \begin{bmatrix}
1 & 0 & 0 \\
0 & 1 & 0 \\
\overline{\beta} & \overline{\alpha} & 1
\end{bmatrix} \begin{bmatrix}
\overline{a} & 0 & 0 \\
0 & \overline{b} & 0 \\
0 & 0 & \overline{c}
\end{bmatrix} \\
\intertext{with an inverse of}
\mathcal{V}^{-1}_{ij} & = \begin{bmatrix}
1 / \overline{a} & 0 & 0 \\
-\overline{\gamma} / \overline{b} & 1 / \overline{b} & 0 \\
-(\overline{\beta} - \overline{\alpha}\overline{\gamma}) / \overline{c} &
-\overline{\alpha} / \overline{c} & 1 / \overline{c}
\end{bmatrix}
\end{align}
\end{subequations}
whose constituents are measured in a coordinate frame with base vectors
\begin{subequations}
    \label{EulerianBaseVectors}
    \begin{align}
    \vec{\mathbfsf{e}}^E_1 & = \boldsymbol{f}^r_1 \bigm/ \overline{a} \\
    \vec{\mathbfsf{e}}^E_2 & = \bigl( \boldsymbol{f}^r_2 - 
    \overline{\gamma} \boldsymbol{f}^r_1 \bigr) \bigm/ \overline{b} \\
    \vec{\mathbfsf{e}}^E_3 & = \bigl( \boldsymbol{f}^r_3 - 
    \overline{\alpha} \boldsymbol{f}^r_2 - 
    ( \overline{\beta} - \overline{\alpha} \overline{\gamma} ) \boldsymbol{f}^r_1 \bigr) \bigm/ \overline{c} 
    \end{align}
\end{subequations}
all of which are described in terms of physical attributes defined as
\begin{equation}
\overline{a} \defeq \mathcal{V}_{11} , \quad
\overline{b} \defeq \mathcal{V}_{22} , \quad
\overline{c} \defeq \mathcal{V}_{33} , \quad
\overline{\alpha} \defeq \frac{\mathcal{V}_{32}}{\mathcal{V}_{22}} , \quad
\overline{\beta} \defeq \frac{\mathcal{V}_{31}}{\mathcal{V}_{11}} , \quad
\overline{\gamma} \defeq \frac{\mathcal{V}_{21}}{\mathcal{V}_{11}}
\label{EulerianPhysicalAttributes}
\end{equation}
where $\overline{a}$, $\overline{b}$ and $\overline{c}$ are elongations, while $\overline{\alpha}$, $\overline{\beta}$ and $\overline{\gamma}$ are magnitudes of shear, i.e., they are the extents of shear at unit elongation.  From conservation of mass, the elongations must be positive ($\overline{a} \in \mathbb{R}_+$, $\overline{b} \in \mathbb{R}_+$, $\overline{c} \in \mathbb{R}_+$), while the shears may be of either sign ($\overline{\alpha} \in \mathbb{R}$, $\overline{\beta} \in \mathbb{R}$, $\overline{\gamma} \in \mathbb{R}$).

\begin{figure}
    \centering
    \includegraphics[width=0.9\linewidth]{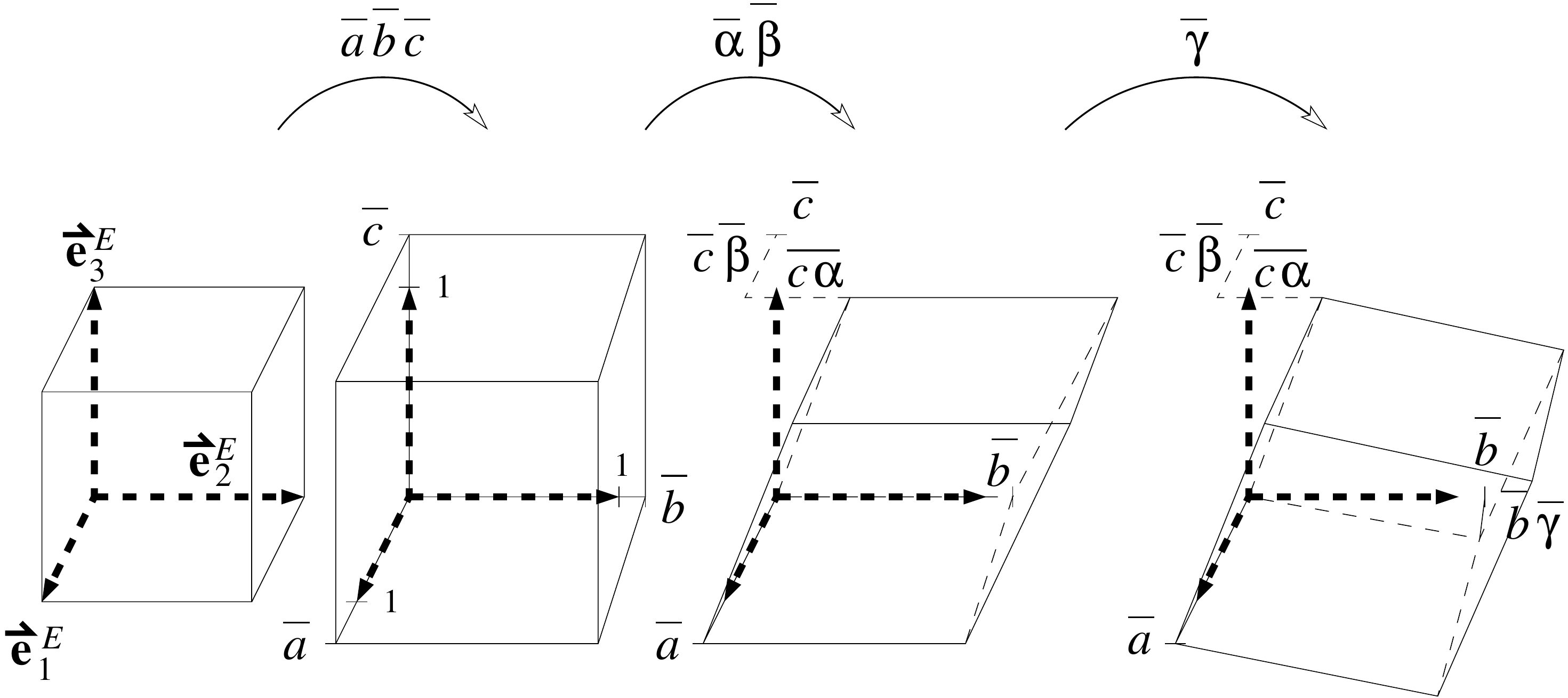}
    \label{fig:EulerianStretch}
    \caption{A geometric interpretation for Eulerian Laplace stretch.}
\end{figure}

According to Eqn.~(\ref{EulerianPhysicalStretch}), the Eulerian Laplace stretch arises from the following sequence of deformations: it starts with three elongations $\overline{a}$, $\overline{b}$ and $\overline{c}$, followed by two out-of-plane shears $\overline{\alpha}$ and $\overline{\beta}$, and then finishes with an in-plane shear $\overline{\gamma}$, as illustrated in Fig.~\ref{fig:EulerianStretch}.  This sequence of deformations is the reverse of that occurring with the Lagrangian Laplace stretch. Two vectors remain invariant under mappings of the Eulerian Laplace stretch, too; they are: vector $\vec{\mathbfsf{e}}^E_1$ establishes the direction of in-plane shear, and vector $\vec{\mathbfsf{e}}^E_1 \times \vec{\mathbfsf{e}}^E_2$ points normal to the plane of in-plane shear.

\section{Examples}

\subsection{Shear-Free Deformations}

Any motion $\boldsymbol{\chi} (\boldsymbol{X},t)$ described by the following deformation gradient quantified in an ortho\-normal coordinate system with base vectors $( \vec{\mathbfsf{e}}_1 ,  \vec{\mathbfsf{e}}_2 , \vec{\mathbfsf{e}}_3 )$ is said to be shear free; specifically,
\begin{equation}
    F_{ij} = 
    \begin{bmatrix}
    \lambda_1 & 0 & 0 \\ 0 & \lambda_2 & 0 \\ 0 & 0 & \lambda_3
    \end{bmatrix}
    \quad \therefore \quad
    B_{ij} = C_{ij} = \begin{bmatrix}
    \lambda_1^2 & 0 & 0 \\ 0 & \lambda_2^2 & 0 \\ 0 & 0 & \lambda_3^2
    \end{bmatrix}
    \label{shearFreeDeformation}
\end{equation}
where $\lambda_1$, $\lambda_2$ and $\lambda_3$ are the three principal stretches that, in this case, obey $\underline{a} = \overline{a} = \lambda_1$, $\underline{b} = \overline{b} = \lambda_2$ and $\underline{c} = \overline{c} = \lambda_3$. The Laplace stretch tensors and their Gram rotations have components of
\begin{equation}
    \mathcal{U}_{ij} = \mathcal{V}_{ij} = \begin{bmatrix}
    \lambda_1 & 0 & 0 \\ 0 & \lambda_2 & 0 \\ 0 & 0 & \lambda_3
    \end{bmatrix}
    \quad \text{with} \quad
    \mathcal{R}^{\!L}_{ij} = \mathcal{R}^{\!E}_{ij} = \begin{bmatrix}
    1 & 0 & 0 \\ 0 & 1 & 0 \\ 0 & 0 & 1
    \end{bmatrix} .
    \label{shearFreeLaplace}
\end{equation}
Consequently, there is no distinction between the triangular Laplace stretches $\boldsymbol{\mathcal{U}}$ and $\boldsymbol{\mathcal{V}}$ and the symmetric polar stretches $\mathbf{U}$ and $\mathbf{V}$ for this class of motions.  The elongations $a$, $b$ and $c$ of Laplace stretch equate with the eigenvalues $\lambda_1$, $\lambda_2$ and $\lambda_3$ of polar stretch.  This relationship between elongations and principal stretches disappears in the presence of shear \cite{Rosakis90}.

\subsection{Pure Shear}

Any motion $\boldsymbol{\chi} ( \boldsymbol{X} , t )$ described by the following deformation gradient quantified in an ortho\-normal coordinate system with base vectors $( \vec{\mathbfsf{e}}_1 ,  \vec{\mathbfsf{e}}_2 , \vec{\mathbfsf{e}}_3 )$ is said to be a pure shear \cite{FreedSrinivasa15}, specifically
\begin{equation}
    F_{ij} = \frac{1}{\sqrt{2}} \begin{bmatrix}
    \sqrt{2} & 0 & 0 \\
    0 & \lambda & \lambda \\ 
    0 & -\lambda^{-1} & \lambda^{-1} 
    \end{bmatrix}
    \label{pureShearDeformationGradient}
\end{equation}
where $\lambda$ is the stretch of pure shear.  This motion is described by Cauchy-Green deformation tensors with components of 
\begin{equation}
    B_{ij} = \begin{bmatrix} 
    1 & 0 & 0 \\ 0 & \lambda^2 & 0 \\ 0 & 0 & \lambda^{-2}
    \end{bmatrix}
    \quad \text{and} \quad
    C_{ij} = \frac{1}{2} \begin{bmatrix}
    2 & 0 & 0 \\
    0 & \lambda^2 + \lambda^{-2} & \lambda^2 - \lambda^{-2} \\
    0 & \lambda^2 - \lambda^{-2} & \lambda^2 + \lambda^{-2} 
    \end{bmatrix}
    \label{pureShearDeformations}
\end{equation}
that produce a Lagrangian Laplace stretch and its Gram rotation of
\begin{subequations}
    \begin{align}
    \mathcal{U}_{ij} & = 
    \frac{1}{\sqrt{ \tfrac{1}{2} ( \lambda^2 + \lambda^{-2} ) }}
    \begin{bmatrix}
    \sqrt{ \tfrac{1}{2} ( \lambda^2 + \lambda^{-2} ) } & 0 & 0 \\
    0 & \tfrac{1}{2} ( \lambda^2 + \lambda^{-2} ) & 
    \tfrac{1}{2} ( \lambda^2 - \lambda^{-2} ) \\
    0 & 0 & 1
    \end{bmatrix} \\
    \intertext{and}
    \mathcal{R}^{\!L}_{ij} & = \frac{1}{\sqrt{\lambda^2 + \lambda^{-2}}} 
    \begin{bmatrix}
    \sqrt{\lambda^2 + \lambda^{-2}} & 0 & 0 \\
    0 & \lambda & \lambda^{-1} \\ 
    0 & -\lambda^{-1} & \lambda 
    \end{bmatrix}
    \end{align}
    \label{pureShearLagLaplace}
\end{subequations}
along with an Eulerian Laplace stretch and its Gram rotation of
\begin{equation}
    \mathcal{V}_{ij} = \begin{bmatrix}
    1 & 0 & 0 \\ 0 & \lambda & 0 \\ 0 & 0 & \lambda^{-1}
    \end{bmatrix}
    \quad \text{and} \quad
    \mathcal{R}^{\!E}_{ij} = \frac{1}{\sqrt{2}} \begin{bmatrix}
    \sqrt{2} & 0 & 0 \\ 0 & 1 & 1 \\ 0 & -1 & 1
    \end{bmatrix}
    \label{pureShearEulLaplace}
\end{equation}
where $\boldsymbol{\mathcal{R}}^{\!E}$ rotates the Eulerian coordinate frame $( \vec{\mathbfsf{e}}_1^E , \vec{\mathbfsf{e}}_2^E , \vec{\mathbfsf{e}}_3^E )$ about the background frame $( \vec{\mathbfsf{e}}_1 , \vec{\mathbfsf{e}}_2 , \vec{\mathbfsf{e}}_3 )$ by a fixed $45^{\circ}$ in the 23~plane; whereas, $\boldsymbol{\mathcal{R}}^{\!L}$ rotates the Lagrangian coordinate frame $( \vec{\mathbfsf{e}}_1^L , \vec{\mathbfsf{e}}_2^L , \vec{\mathbfsf{e}}_3^L )$ from the Eulerian frame  $( \vec{\mathbfsf{e}}_1^E , \vec{\mathbfsf{e}}_2^E , \vec{\mathbfsf{e}}_3^E )$ at $\lambda = 1$ towards the background frame $( \vec{\mathbfsf{e}}_1 , \vec{\mathbfsf{e}}_2 , \vec{\mathbfsf{e}}_3 )$ as $\lambda \to \infty$.  

The above components for Eulerian Laplace stretch $\mathcal{V}_{ij}$ support Lodge's statement that pure shear is not a shearing deformation; it is a shear-free deformation in disguise \cite{Lodge64,Lodge74}.  Lodge justifies this position by pointing out that the eigen\-vectors for stretch do not rotate in a body during pure shears like they do during simple shears.  

Here the elongations relate as $\underline{a} = \overline{a} = 1$ while $\underline{b} = \sqrt{(\lambda^2 + \lambda^{-2})/2}$ and $\overline{b} = \lambda$ with $\underline{c} = 1 / \sqrt{(\lambda^2 + \lambda^{-2})/2}$ and $\overline{c} = \lambda^{-1}$, whereas the shears relate as $\underline{\alpha} = ( \lambda^2 - \lambda^{-2} ) / ( \lambda^2 + \lambda^{-2} )$ and $\overline{\alpha} = 0$ with $\underline{\beta} = \overline{\beta} = \underline{\gamma} = \overline{\gamma} = 0$.

\subsection{Simple Shear}

Any motion $\boldsymbol{\chi} ( \boldsymbol{X},t )$ described by the following deformation gradient quantified in an ortho\-normal coordinate system with base vectors $( \vec{\mathbfsf{e}}_1 ,  \vec{\mathbfsf{e}}_2 , \vec{\mathbfsf{e}}_3 )$ constitutes a shearing motion; specifically,
\begin{equation}
F_{ij} = \begin{bmatrix}
1 & 0 & \beta \\ 0 & 1 & 0 \\ 0 & 0 & 1 
\end{bmatrix}
\label{outOfPlaneDeformationGradient1}
\end{equation}
whose Cauchy-Green deformation tensors have components of
\begin{equation}
B_{ij} = \begin{bmatrix}
1 + \beta^2 & 0 & \beta \\
0 & 1 & 0 \\
\beta & 0 & 1
\end{bmatrix} 
\quad \text{and} \quad
C_{ij} = \begin{bmatrix}
1 & 0 & \beta \\
0 & 1 & 0 \\
\beta & 0 & 1 + \beta^2
\end{bmatrix}
\label{outOfPlaneDeformations1}
\end{equation}
with its Lagrangian Laplace stretch and rotation having components of
\begin{equation}
\mathcal{U}_{ij} = \begin{bmatrix}
1 & 0 & \beta \\
0 & 1 & 0 \\
0 & 0 & 1
\end{bmatrix}
\quad \text{and} \quad
\mathcal{R}^{\!L}_{ij} = \begin{bmatrix}
1 & 0 & 0 \\ 0 & 1 & 0 \\ 0 & 0 & 1
\end{bmatrix}
\label{outOfPlaneLagrangian1}
\end{equation}
along with its Eulerian Laplace stretch and rotation having components of
\begin{subequations}
    \label{outOfPlaneEulerian1}
    \begin{align}
    \mathcal{V}_{ij} & = \begin{bmatrix}
    \sqrt{1 + \beta^2} & 0 & 0 \\
    0 & 1 & 0 \\
    \beta / \sqrt{1 + \beta^2} & 0 & 1 / \sqrt{1 + \beta^2}
    \end{bmatrix} \\
    \intertext{and}
    \mathcal{R}^{\!E}_{ij} & = \begin{bmatrix}
    1 / \sqrt{1+\beta^2} & 0 & \beta / \sqrt{1 + \beta^2} \\
    0 & 1 & 0 \\
    -\beta / \sqrt{1 + \beta^2} & 0 & 1 / \sqrt{1 + \beta^2}
    \end{bmatrix}
    \end{align}
\end{subequations}
with the Eulerian Laplace stretch $\mathcal{V}_{ij}$ having diagonal elements akin to those of pure shear (cf.\ Eqn.~\ref{pureShearEulLaplace}), plus an off-diagonal simple shearing that is attenuated by the extent of pure shearing present. 

From a rheometric viewpoint, making stress a function of the Eulerian Laplace stretch would enable first- and second-normal stress differences to occur, with the first exceeding the second in magnitude, and they being of opposite sign.  A Weisenberg effect would occur, because of a compressive stretch that would set up in the hoop direction.  Furthermore, the shear stress would thin, because of an effect that $\sqrt{1 + \gamma^2}$ would have on the shear strain $\gamma / \sqrt{1 + \gamma^2}$.  All of these `effects' occur in polymeric liquids \cite{Birdetal87}. 

Here the elongations relate as $\underline{a} = 1$ and $\overline{a} = \sqrt{1 + \beta^2}$ while $\underline{b} = \overline{b} = 1$ with $\underline{c} = 1$ and $\overline{c} = 1 / \sqrt{1 + \beta^2}$, whereas the shears relate as $\underline{\beta} = \beta$ and $\overline{\beta} = \beta / (1 + \beta^2)$ with $\underline{\alpha} = \overline{\alpha} = \underline{\gamma} = \overline{\gamma} = 0$.

\section{Frameworks for Constitutive Development}

A time rate-of-change in the work being done at a particle by tractions applied to its body results in a source for internal power caused by stresses, often evaluated per unit mass.  Here we construct sets of thermo\-dynamic conjugate pairs for both the Lagrangian and Eulerian frameworks when using Laplace stretch as one's kinematic variable.  The constituents of these pairs relate to one another via constitutive equations (a topic for future papers).  To facilitate such endeavors, bijective maps are derived that convert stress and velocity-gradient tensor components into their associated thermo\-dynamic stresses and strain rates, the latter of which are scalar fields.

\subsection{Lagrangian Stress-Strain Attributes}

In terms of Lagrangian fields, stress power $\dot{W}$ can be written as $\tfrac{1}{\rho_0} \mathrm{tr} ( \mathbf{S} \dot{\mathbf{E}} )$ wherein $\mathbf{S}$ is the second Piola-Kirchhoff stress, $\mathbf{E} \defeq \tfrac{1}{2} ( \mathbf{C} - \mathbf{I} )$ is the Green strain, and $\rho_0$ is the initial mass density at a particle of interest in a body. 

It is easily verified that
\begin{equation}
\dot{W} = \tfrac{1}{\rho_0} \mathrm{tr} ( \mathbf{S} \dot{\mathbf{E}} ) = 
\tfrac{1}{\rho_0} \mathrm{tr} ( \boldsymbol{\mathcal{S}} 
\boldsymbol{\mathcal{L}}\mbox{}^L ) 
\quad \text{where} \quad
\boldsymbol{\mathcal{S}} \defeq \boldsymbol{\mathcal{U}} \mathbf{S} \, \boldsymbol{\mathcal{U}}^{\mathsf{T}} ,
\quad
\boldsymbol{\mathcal{L}}\mbox{}^L \defeq \dot{\boldsymbol{\mathcal{U}}} \boldsymbol{\mathcal{U}}^{-1}
\label{LagrangianStressPower}
\end{equation}
given that $\mathbf{F} = \boldsymbol{\mathcal{R}}^{\!L} \boldsymbol{\mathcal{U}}$.  The Lagrangian stress $\boldsymbol{\mathcal{S}}$ is symmetric because the second Piola-Kirchhoff stress $\mathbf{S}$ is symmetric, and the Lagrangian velocity gradient $\boldsymbol{\mathcal{L}}\mbox{}^L$ is upper-triangular---a consequence of the group that stretch $\boldsymbol{\mathcal{U}}$ belongs to.  The above expression for stress power reduces to a sum of six scalar contributions; specifically
\begin{equation}
\rho_0 \dot{W} = \mathcal{S}_{11} \mathcal{L}^L_{11} + 
\mathcal{S}_{21} \mathcal{L}^L_{12} + \mathcal{S}_{22} \mathcal{L}^L_{22} +
\mathcal{S}_{31} \mathcal{L}^L_{13} + \mathcal{S}_{32} \mathcal{L}^L_{23} + 
\mathcal{S}_{33} \mathcal{L}^L_{33}
\label{LagrangianWorkRate}
\end{equation}
wherein
\begin{equation}
\mathcal{L}^L_{ij} = \dot{\mathcal{U}}_{ik\,} \mathcal{U}_{kj}^{-1} = 
\begin{bmatrix}
\dot{\underline{a}} / \underline{a} & \underline{a} \dot{\underline{\gamma}} /
\underline{b} & \underline{a} ( \dot{\underline{\beta}} - \underline{\alpha} 
\dot{\underline{\gamma}} ) / \underline{c} \\
0 & \dot{\underline{b}} / \underline{b} & \underline{b} 
\dot{\underline{\alpha}} / \underline{c} \\
0 & 0 & \dot{\underline{c}} / \underline{c}
\end{bmatrix}
\label{LagrangianVelocityGradient}
\end{equation}
and we observe that the diagonal rates are logarithmic, while the off-diagonal rates are not logarithmic.  (A very different triangular velocity gradient, viz., Eqn.~(\ref{EulerianVelocityGradient}), arises in the Eulerian construction that follows.)  How to construct proper finite differences to approximate derivatives for the physical attributes of Laplace stretch is discussed in Ref.~\cite{FreedZamani18}.

Expressing Eqn.~(\ref{LagrangianWorkRate}) in terms of thermo\-dynamic conjugate pairs is not a unique process, cf.\ Ref.~\cite{Freed17}.  Here we shall consider a pairing described by 
\begin{equation}
\rho_0 \dot{W} = \underline{\pi} \dot{\underline{\delta}} + \sum_{i=1}^3 
\bigl( \underline{\sigma}_i \dot{\underline{\varepsilon}}_i + 
\underline{\tau}_i \dot{\underline{\gamma}}_i \bigr)
\end{equation}
whose seven, conjugate, stress-strain pairs are defined as follows: a uniform bulk response is governed by a Lagrangian pressure $\underline{\pi}$ and a Lagrangian dilatation $\underline{\delta}$ defined by
\begin{subequations}
    \label{LagrangianConjugatePairs}
    \begin{align}
    \underline{\pi} & \defeq \mathcal{S}_{11} + \mathcal{S}_{22} + \mathcal{S}_{33} &
    \underline{\delta} & \defeq \ln \sqrt[3]{\frac{\underline{a}}{\underline{a}_0}
        \frac{\underline{b}}{\underline{b}_0} \frac{\underline{c}}{\underline{c}_0}} & 
    \dot{\underline{\delta}} & = \frac{1}{3} \left( \frac{\dot{\underline{a}}}
    {\underline{a}} + \frac{\dot{\underline{b}}}{\underline{b}} + 
    \frac{\dot{\underline{c}}}{\underline{c}} \right) \\
    \intertext{while the squeeze (pure shear) responses are governed by Lagrangian normal-stress differences $\underline{\sigma}_i$ and Lagrangian squeezes $\underline{\varepsilon}_i$ defined by}
    \underline{\sigma}_1 & \defeq \mathcal{S}_{11} - \mathcal{S}_{22} & 
    \underline{\varepsilon}_1 & \defeq \ln \sqrt[3]{\frac{\underline{a}}
        {\underline{a}_0} \frac{\underline{b}_0}{\underline{b}}} & \dot{\underline{\varepsilon}}_1 & = 
    \frac{1}{3} \left( \frac{\dot{\underline{a}}}{\underline{a}} - 
    \frac{\dot{\underline{b}}}{\underline{b}} \right) \\
    \underline{\sigma}_2 & \defeq \mathcal{S}_{22} - \mathcal{S}_{33} & 
    \underline{\varepsilon}_2 & \defeq \ln \sqrt[3]{\frac{\underline{b}}
        {\underline{b}_0} \frac{\underline{c}_0}{\underline{c}}} & \dot{\underline{\varepsilon}}_2 & = 
    \frac{1}{3} \left( \frac{\dot{\underline{b}}}{\underline{b}} - 
    \frac{\dot{\underline{c}}}{\underline{c}} \right) \\
    \underline{\sigma}_3 & \defeq \mathcal{S}_{33} - \mathcal{S}_{11} & 
    \underline{\varepsilon}_3 & \defeq \ln \sqrt[3]{\frac{\underline{c}}
        {\underline{c}_0} \frac{\underline{a}_0}{\underline{a}}} & \dot{\underline{\varepsilon}}_3 & = 
    \frac{1}{3} \left( \frac{\dot{\underline{c}}}{\underline{c}} - 
    \frac{\dot{\underline{a}}}{\underline{a}} \right) \\
    \intertext{of which two are independent because $\underline{\sigma}_3 = -( \underline{\sigma}_1 + \underline{\sigma}_2 )$ and $\underline{\varepsilon}_3 = -( \underline{\varepsilon}_1 + \underline{\varepsilon}_2 )$, while the (simple) shear responses are governed by Lagrangian shear stresses $\underline{\tau}_i$ and Lagrangian shear strains $\underline{\gamma}_i$ defined by}
    \underline{\tau}_1 & \defeq \frac{\underline{b}}{\underline{c}} \, \mathcal{S}_{32} &
    \underline{\gamma}_1 & \defeq \underline{\alpha} - \underline{\alpha}_0 & 
    \dot{\underline{\gamma}}_1 & = \dot{\underline{\alpha}} \\
    \underline{\tau}_2 & \defeq \frac{\underline{a}}{\underline{c}} \, \mathcal{S}_{31} & 
    \underline{\gamma}_2 & \defeq \underline{\beta} - \underline{\beta}_0 & 
    \dot{\underline{\gamma}}_2 & = \dot{\underline{\beta}} \\
    \underline{\tau}_3 & \defeq \frac{\underline{a}}{\underline{b}} \, \mathcal{S}_{21} -
    \frac{\underline{a} \underline{\alpha}}{\underline{c}} \, \mathcal{S}_{31} &  
    \underline{\gamma}_3 & \defeq \underline{\gamma} - \underline{\gamma}_0 & 
    \dot{\underline{\gamma}}_3 & = \dot{\underline{\gamma}} 
    \end{align}
\end{subequations}
wherein $\underline{a}_0$, $\underline{b}_0$ and $\underline{c}_0$ are their initial elongation ratios, and where $\underline{\alpha}_0$, $\underline{\beta}_0$ and $\underline{\gamma}_0$ are their initial shears.

Bijective maps exist to transform tensor components into thermo\-dynamic stress--strain-rate attributes that, for isotropic materials,\footnote{
    See Ref.~\cite{Freed17} for one way to extend this approach to anisotropic materials.
}
are described by
\begin{subequations}
    \label{LagrangianMaps}
    \begin{align}
    \left\{ \begin{matrix}
    \underline{\pi} \\ \underline{\sigma}_1 \\ \underline{\sigma}_2 \\
    \underline{\tau}_1 \\ \underline{\tau}_2 \\ \underline{\tau}_3
    \end{matrix} \right\} & = \begin{bmatrix}
    1 & 1 & 1 & 0 & 0 & 0 \\
    1 & -1 & 0 & 0 & 0 & 0 \\
    0 & 1 & -1 & 0 & 0 & 0 \\
    0 & 0 & 0 & \underline{b} / \underline{c} & 0 & 0 \\
    0 & 0 & 0 & 0 & \underline{a} / \underline{c} & 0 \\
    0 & 0 & 0 & 0 & -\underline{a} \underline{\alpha} / \underline{c} & 
    \underline{a} / \underline{b}
    \end{bmatrix} \left\{ \begin{matrix}
    \mathcal{S}_{11} \\ \mathcal{S}_{22} \\ \mathcal{S}_{33} \\
    \mathcal{S}_{32} \\ \mathcal{S}_{31} \\ \mathcal{S}_{21}
    \end{matrix} \right\} \\
    \intertext{with $\underline{\sigma}_3 = -\underline{\sigma}_1 - \underline{\sigma}_2$, and}
    \left\{ \begin{matrix}
    \dot{\underline{\delta}} \\ \dot{\underline{\varepsilon}}_1 \\
    \dot{\underline{\varepsilon}}_2 \\ \dot{\underline{\gamma}}_1 \\
    \dot{\underline{\gamma}}_2 \\ \dot{\underline{\gamma}}_3
    \end{matrix} \right\} & = \begin{bmatrix}
    1/3 & 1/3 & 1/3 & 0 & 0 & 0 \\
    1/3 & -1/3 & 0 & 0 & 0 & 0 \\
    0 & 1/3 & -1/3 & 0 & 0 & 0 \\
    0 & 0 & 0 & \underline{c} / \underline{b} & 0 & 0 \\
    0 & 0 & 0 & 0 & \underline{c} / \underline{a} & 
    \underline{b} \underline{\alpha} / \underline{a} \\
    0 & 0 & 0 & 0 & 0 & \underline{b} / \underline{a}
    \end{bmatrix} \left\{ \begin{matrix}
    \mathcal{L}^L_{11} \\ \mathcal{L}^L_{22} \\ \mathcal{L}^L_{33} \\
    \mathcal{L}^L_{23} \\ \mathcal{L}^L_{13} \\ \mathcal{L}^L_{12}
    \end{matrix} \right\}
    \end{align}
\end{subequations}
with $\dot{\underline{\varepsilon}}_3 = -\dot{\underline{\varepsilon}}_1 - \dot{\underline{\varepsilon}}_2$.  

These strain-rate attributes can be integrated to get the Lagrangian thermo\-dynamic strains $\underline{\delta}$, $\underline{\varepsilon}_1$, $\underline{\varepsilon}_2$, $\underline{\varepsilon}_3$, $\underline{\gamma}_1$, $\underline{\gamma}_2$ and $\underline{\gamma}_3$ by choosing initial conditions of  $\underline{\delta} |_0 = \underline{\varepsilon}_1 |_0 = \underline{\varepsilon}_2 |_0 = \underline{\varepsilon}_3 |_0 = \underline{\gamma}_1 |_0 = \underline{\gamma}_2 |_0 = \underline{\gamma}_3 |_0 = 0$ provided that the initial elongation ratios have been specified as $\underline{a}_0$, $\underline{b}_0$ and $\underline{c}_0$ and that the initial magnitudes of shear have been specified as $\underline{\alpha}_0$, $\underline{\beta}_0$ and $\underline{\gamma}_0$.

At this juncture, constitutive equations between stress-strain attributes of the thermo\-dynamic conjugate pairs $( \underline{\pi} , \underline{\delta} )$, $( \underline{\sigma}_1 , \underline{\varepsilon}_1 )$, $( \underline{\sigma}_2 , \underline{\varepsilon}_2 )$, $( \underline{\tau}_1 , \underline{\gamma}_1 )$, $( \underline{\tau}_2 , \underline{\gamma}_2 )$ and $( \underline{\tau}_3 , \underline{\gamma}_3 )$ are to be introduced (a topic for future works) to solve for the Lagrangian thermo\-dynamic stresses $\underline{\pi}$, $\underline{\sigma}_1$, $\underline{\sigma}_2$, $\underline{\sigma}_3$, $\underline{\tau}_1$, $\underline{\tau}_2$ and $\underline{\tau}_3$. These updated stress attributes map into our Lagrangian stress components $\mathcal{S}_{ij}$ as
\begin{equation}
\left\{ \begin{matrix}
\mathcal{S}_{11} \\ \mathcal{S}_{22} \\ \mathcal{S}_{33} \\
\mathcal{S}_{23} = \mathcal{S}_{32} \\ \mathcal{S}_{13} = \mathcal{S}_{31} \\ \mathcal{S}_{12} = \mathcal{S}_{21}
\end{matrix} \right\} = \begin{bmatrix}
1/3 & 2/3 & 1/3 & 0 & 0 & 0 \\
1/3 & -1/3 & 1/3 & 0 & 0 & 0 \\
1/3 & -1/3 & -2/3 & 0 & 0 & 0 \\
0 & 0 & 0 & \underline{c} / \underline{b} & 0 & 0 \\
0 & 0 & 0 & 0 & \underline{c} / \underline{a} & 0 \\
0 & 0 & 0 & 0 & \underline{b} \underline{\alpha} / \underline{a} & 
\underline{b} / \underline{a}
\end{bmatrix} \left\{ \begin{matrix}
\underline{\pi} \\ \underline{\sigma}_1 \\ \underline{\sigma}_2 \\
\underline{\tau}_1 \\ \underline{\tau}_2 \\ \underline{\tau}_3
\end{matrix} \right\}
\end{equation}
from which the second Piola-Kirchhoff stress $\mathbf{S} = S_{ij} \, \vec{\mathbfsf{e}}_i \otimes \vec{\mathbfsf{e}}_j$ is retrieved via $\mathbf{S} = \boldsymbol{\mathcal{U}}^{-1} \boldsymbol{\mathcal{S}} \boldsymbol{\mathcal{U}}^{-\mathsf{T}}$, i.e., $S_{ij} = \mathcal{U}^{-1}_{ik} \mathcal{S}_{k\ell\,} \mathcal{U}^{-1}_{j\ell}$, and from here any commonly used stress tensor can be gotten.

Although $\underline{\sigma}_3$ and $\dot{\underline{\varepsilon}}_3$ are not needed from a constitutive perspective, they are required to correctly calculate stress power.

\subsection{Eulerian Stress-Strain Attributes}

In terms of Eulerian fields, stress power $\dot{W}$ can be written as $\tfrac{1}{\rho_0} \mathrm{tr} ( \boldsymbol{\tau} \mathbf{D} )$ wherein $\boldsymbol{\tau} = \mathbf{F} \mathbf{S} \mathbf{F}^{\mathsf{T}}$ is the Kirchhoff stress, which relates to Cauchy stress $\mathbf{T}$ via $\boldsymbol{\tau} \defeq \det ( \mathbf{F} ) \mathbf{T} = \tfrac{\rho_0}{\rho} \mathbf{T}$, and where $\mathbf{D} \defeq \tfrac{1}{2} ( \mathbf{L} + \mathbf{L}^{\mathsf{T}} ) = \mathbf{F}^{-\mathsf{T}} \dot{\mathbf{E}} \mathbf{F}^{-1}$ is the symmetric part of velocity gradient $\mathbf{L} \defeq \dot{\mathbf{F}} \mathbf{F}^{-1}$, with $\rho$ being the current mass density.

It can be shown that
\begin{subequations}
    \label{EulerianStressPower}
    \begin{align}
    \dot{W} & = \tfrac{1}{\rho_0} \mathrm{tr} ( \boldsymbol{\tau} \mathbf{D} ) = 
    \tfrac{1}{\rho_0} \mathrm{tr} \bigl( \boldsymbol{\tau} \boldsymbol{\mathcal{L}}^E \bigr) \\
    \intertext{given that $\mathbf{F} = \boldsymbol{\mathcal{VR}}^{\!E}$, where this Eulerian velocity gradient $\boldsymbol{\mathcal{L}}^E$ is defined by}
    \boldsymbol{\mathcal{L}}^E & \defeq 
    \overset{\circ}{\boldsymbol{\mathcal{V}}} \boldsymbol{\mathcal{V}}^{-1}
    \quad \text{wherein} \quad
    \overset{\circ}{\boldsymbol{\mathcal{V}}} \defeq 
    \dot{\boldsymbol{\mathcal{V}}} + \boldsymbol{\mathcal{V}} \boldsymbol{\mathit{\Omega}}^E - \boldsymbol{\mathit{\Omega}}^E \boldsymbol{\mathcal{V}}
    \end{align}
\end{subequations}
with $\overset{\circ}{\boldsymbol{\mathcal{V}}}$ being an objective co-rotational derivative for this measure of stretch, and $\boldsymbol{\mathit{\Omega}}^E \defeq \dot{\boldsymbol{\mathcal{R}}}\mbox{}^{\!E} \boldsymbol{\mathcal{R}}^{\!E^{\mathsf{T}}}$ being a spin of an Eulerian co\-ordinate axes $( \vec{\mathbfsf{e}}_1^E , \vec{\mathbfsf{e}}_2^E , \vec{\mathbfsf{e}}_3^E )$ about the reference axes $( \vec{\mathbfsf{e}}_1 , \vec{\mathbfsf{e}}_2 , \vec{\mathbfsf{e}}_3 )$.  

Consequently, stress power $\rho_0 \dot{W} = \mathrm{tr} \bigl( \boldsymbol{\tau} \boldsymbol{\mathcal{L}}^E \bigr)$ arises from two sources in this Eulerian construction, viz. $\dot{W} = \dot{W}_1 + \dot{W}_2$.  The first is energetic, i.e.,
\begin{subequations}
    \begin{align}
    \dot{W}_1 & \defeq \tfrac{1}{\rho_0} \mathrm{tr} \bigl( \boldsymbol{\tau} \dot{\boldsymbol{\mathcal{V}}} \boldsymbol{\mathcal{V}}^{-1} \bigr) 
    \label{EulerianStressPower1} \\
    \intertext{while the second satifies objectivity, viz.,}
    \dot{W}_2 & \defeq \tfrac{1}{\rho_0} \mathrm{tr} \bigl( \boldsymbol{\tau} \boldsymbol{\mathcal{V}} \boldsymbol{\mathit{\Omega}}^E \boldsymbol{\mathcal{V}}^{-1} \bigr)
    \label{EulerianStressPower2}
    \end{align}
\end{subequations}
noting that $\mathrm{tr} ( \boldsymbol{\tau} \boldsymbol{\mathit{\Omega}}^E ) = 0$.  Thermo\-dynamic stress-strain conjugate pairs can be established in terms of the energetic expression (\ref{EulerianStressPower1}).  The objective correction (\ref{EulerianStressPower2}) is required to quantify the work being done, but it plays no role when creating our Eulerian stress-strain attributes, as every term in this sum has a component of spin in it; therefore, $\dot{W}_2 = 0$ whenever $\boldsymbol{\mathit{\Omega}}^E = \mathbf{0}$.

Because $\dot{\boldsymbol{\mathcal{V}}} \boldsymbol{\mathcal{V}}^{-1} = \dot{\mathcal{V}}_{ik} \mathcal{V}_{kj}^{-1} \, \vec{\mathbfsf{e}}_i \otimes \vec{\mathbfsf{e}}_j$ has components that are lower triangular, a consequence of the group that tensor $\boldsymbol{\mathcal{V}}$ belongs to, the first contribution to stress power put forward in Eqn.~(\ref{EulerianStressPower1}) reduces to a sum of six scalar contributions; specifically,
\begin{multline}
\rho_0 \dot{W}_1 = 
\tau_{11} \dot{\mathcal{V}}_{1i} \mathcal{V}^{-1}_{i1} +
\tau_{12} \dot{\mathcal{V}}_{2i} \mathcal{V}^{-1}_{i1} + 
\tau_{13} \dot{\mathcal{V}}_{3i} \mathcal{V}^{-1}_{i1} \\ +
\tau_{22} \dot{\mathcal{V}}_{2i} \mathcal{V}^{-1}_{i2} + 
\tau_{23} \dot{\mathcal{V}}_{3i} \mathcal{V}^{-1}_{i2} + 
\tau_{33} \dot{\mathcal{V}}_{3i} \mathcal{V}^{-1}_{i3}
\label{EulerianWorkRate}
\end{multline}
wherein
\begin{equation}
\label{EulerianVelocityGradient}
\dot{\mathcal{V}}_{ik} \mathcal{V}^{-1}_{kj} = \begin{bmatrix}
\frac{\dot{\overline{a}}}{\overline{a}} & 0 & 0 \\
\dot{\overline{\gamma}} + \overline{\gamma} \left( 
\frac{\dot{\overline{a}}}{\overline{a}} -
\frac{\dot{\overline{b}}}{\overline{b}} \right) & 
\frac{\dot{\overline{b}}}{\overline{b}} & 0 \\
\dot{\overline{\beta}} - \overline{\gamma} \, \dot{\overline{\alpha}} + 
\overline{\beta} \left( \frac{\dot{\overline{a}}}{\overline{a}} - 
\frac{\dot{\overline{c}}}{\overline{c}} \right) - 
\overline{\alpha} \overline{\gamma} \left( 
\frac{\dot{\overline{b}}}{\overline{b}} - 
\frac{\dot{\overline{c}}}{\overline{c}} \right) & 
\dot{\overline{\alpha}} + \overline{\alpha} \left( 
\frac{\dot{\overline{b}}}{\overline{b}} - 
\frac{\dot{\overline{c}}}{\overline{c}} \right) & 
\frac{\dot{\overline{c}}}{\overline{c}}   
\end{bmatrix}
\end{equation}
which is strikingly different from that of its Lagrangian counterpart $\dot{\boldsymbol{\mathcal{U}}} \boldsymbol{\mathcal{U}}^{-1}$ found in Eqn.~(\ref{LagrangianVelocityGradient}).\footnote{
   Curiously, $\boldsymbol{\mathcal{U}}^{-1} \dot{\boldsymbol{\mathcal{U}}}$ has components akin to Eqn.~(\ref{EulerianVelocityGradient}), except its components are upper triangular instead of lower triangular, and are expressed in terms of the Lagrangian stretch attributes instead of their Eulerian counterparts.
}
Present here are the squeeze rates $\dot{\overline{\varepsilon}}_1 = \tfrac{1}{3} \bigl( \dot{\overline{a}} / \overline{a} - \dot{\overline{b}} / \overline{b} \bigr)$, etc., which appear in the off-diagonal terms, along with their corresponding shear rates, e.g., $\dot{\overline{\gamma}}$, thereby substantiating our assumed construction of conjugate pairs.

In Eqn.~(\ref{EulerianVelocityGradient}), a clear delineation exists between pure and simple shearing deformations.  Such a delineation does not arise whenever one uses symmetric measures for stretch, where an isotropic-deviatoric decomposition is the extent to which such fields can be deconstructed.

Expressing Eqn.~(\ref{EulerianWorkRate}) in terms of Eulerian, thermo\-dynamic, conjugate pairs, analogous to those considered for the Lagrangian frame, one can write
\begin{equation}
\rho_0 \dot{W}_1 = \overline{\pi} \dot{\overline{\delta}} + \sum_{i=1}^3 
\bigl( \overline{\sigma}_i \dot{\overline{\varepsilon}}_i + 
\overline{\tau}_i \dot{\overline{\gamma}}_i \bigr)
\label{EulerianStressWorking}
\end{equation}
whose seven, conjugate, stress-strain pairs are defined as follows: a uniform bulk response is governed by an Eulerian pressure $\overline{\pi}$ and an Eulerian dilatation $\overline{\delta}$ defined by
\begin{subequations}
    \label{EulerianConjugatePairs}
    \begin{align}
    \overline{\pi} & \defeq \tau_{11} + \tau_{22} + \tau_{33} &
    \overline{\delta} & \defeq \ln \sqrt[3]{\frac{\overline{a}}{\overline{a}_0}
        \frac{\overline{b}}{\overline{b}_0} \frac{\overline{c}}{\overline{c}_0}} & 
    \dot{\overline{\delta}} & = \frac{1}{3} \left( \frac{\dot{\overline{a}}}
    {\overline{a}} + \frac{\dot{\overline{b}}}{\overline{b}} + 
    \frac{\dot{\overline{c}}}{\overline{c}} \right) \\
    \intertext{while the squeeze (pure shear) responses are governed by Eulerian normal-stress differences $\overline{\sigma}_i$ and Eulerian squeezes $\overline{\varepsilon}_i$ defined by}
    \overline{\sigma}_1 & \defeq \tau_{11} - \tau_{22} + 
    3 \overline{\gamma} \tau_{12} &
    \overline{\varepsilon}_1 & \defeq \ln \sqrt[3]{\frac{\overline{a}}
        {\overline{a}_0} \frac{\overline{b}_0}{\overline{b}}} & \dot{\overline{\varepsilon}}_1 & = 
    \frac{1}{3} \left( \frac{\dot{\overline{a}}}{\overline{a}} - 
    \frac{\dot{\overline{b}}}{\overline{b}} \right) \\
    \overline{\sigma}_2 & \defeq \left\{
    \begin{aligned} 
    \mbox{} & \tau_{22} - \tau_{33} \\ \mbox{} & + 
    3 \overline{\alpha} ( \tau_{23} - \overline{\gamma} \tau_{13} )
    \end{aligned} \right. & 
    \overline{\varepsilon}_2 & \defeq \ln \sqrt[3]{\frac{\overline{b}}
        {\overline{b}_0} \frac{\overline{c}_0}{\overline{c}}} & \dot{\overline{\varepsilon}}_2 & = 
    \frac{1}{3} \left( \frac{\dot{\overline{b}}}{\overline{b}} - 
    \frac{\dot{\overline{c}}}{\overline{c}} \right) \\
    \overline{\sigma}_3 & \defeq -\tau_{11} + \tau_{33} - 
    3 \overline{\beta} \tau_{13} & 
    \overline{\varepsilon}_3 & \defeq \ln \sqrt[3]{\frac{\overline{c}}
        {\overline{c}_0} \frac{\overline{a}_0}{\overline{a}}} & \dot{\overline{\varepsilon}}_3 & = 
    \frac{1}{3} \left( \frac{\dot{\overline{c}}}{\overline{c}} - 
    \frac{\dot{\overline{a}}}{\overline{a}} \right) \\
    \intertext{of which only two are independent, while the (simple) shear responses are governed by Eulerian shear stresses $\overline{\tau}_i$ and strains $\overline{\gamma}_i$ defined by}
    \overline{\tau}_1 & \defeq \tau_{23} - \overline{\gamma} \tau_{13} &
    \overline{\gamma}_1 & \defeq \overline{\alpha} - \overline{\alpha}_0 & 
    \dot{\overline{\gamma}}_1 & = \dot{\overline{\alpha}} \\
    \overline{\tau}_2 & \defeq \tau_{13} & 
    \overline{\gamma}_2 & \defeq \overline{\beta} - \overline{\beta}_0 & 
    \dot{\overline{\gamma}}_2 & = \dot{\overline{\beta}} \\
    \overline{\tau}_3 & \defeq \tau_{12} &  
    \overline{\gamma}_3 & \defeq \overline{\gamma} - \overline{\gamma}_0 & 
    \dot{\overline{\gamma}}_3 & = \dot{\overline{\gamma}} 
    \end{align}
\end{subequations}
wherein $\overline{a}_0$, $\overline{b}_0$ and $\overline{c}_0$ are their initial elongation ratios, and where $\overline{\alpha}_0$, $\overline{\beta}_0$ and $\overline{\gamma}_0$ are their initial shear offsets.  

The sets of thermo\-dynamic conjugate pairs for the Lagrangian and Eulerian frameworks are taken to be the same.  Each set is composed of three modes: one pair to describe uniform dilatation, three pairs to describe pure shears, and three pairs to describe simple shears.  In both cases, only two of the three pure-shear pairs are independent, thereby resulting in sets of six, independent, conjugate pairs that have direct connections with the six independent components of stress and stretch rate.

Bijective maps exist to transform tensor components into thermo\-dynamic stress--strain-rate attributes that, for isotropic materials, are described by
\begin{subequations}
    \label{EulerianMaps}
    \begin{align}
    \left\{ \begin{matrix}
    \overline{\pi} \\ \overline{\sigma}_1 \\ \overline{\sigma}_2 \\
    \overline{\tau}_1 \\ \overline{\tau}_2 \\ \overline{\tau}_3
    \end{matrix} \right\} & = \begin{bmatrix}
    1 & 1 & 1 & 0 & 0 & 0 \\
    1 & -1 & 0 & 0 & 0 & 3 \overline{\gamma} \\
    0 & 1 & -1 & 3 \overline{\alpha} & -3 \overline{\alpha} \overline{\gamma} & 0 \\
    0 & 0 & 0 & 1 & -\overline{\gamma} & 0 \\
    0 & 0 & 0 & 0 & 1 & 0 \\
    0 & 0 & 0 & 0 & 0 & 1
    \end{bmatrix} \left\{ \begin{matrix}
    \tau_{11} \\ \tau_{22} \\ \tau_{33} \\
    \tau_{32} \\ \tau_{31} \\ \tau_{21}
    \end{matrix} \right\} \\
    \intertext{with}
    \overline{\sigma}_3 & = -\overline{\sigma}_1 - \overline{\sigma}_2 + 
    3 \bigl( \overline{\alpha} \overline{\tau}_1 - 
    \overline{\beta} \overline{\tau}_2 + 
    \overline{\gamma} \overline{\tau}_3 \bigr)  \\ 
    \intertext{which arises from the constraint equation}
    \left\{ \begin{matrix}
    \overline{\sigma}_1 - 3 \overline{\gamma} \overline{\tau}_3 \\
    \overline{\sigma}_2 - 3 \overline{\alpha} \overline{\tau}_1 \\
    \overline{\sigma}_3 + 3 \overline{\beta} \overline{\tau}_2
    \end{matrix} \right\} & = \begin{bmatrix}
    1 & -1 & 0 \\ 0 & 1 & -1 \\ -1 & 0 & 1
    \end{bmatrix} \left\{ \begin{matrix}
    \tau_{11} \\ \tau_{22} \\ \tau_{33}
    \end{matrix} \right\}
    \notag
    \intertext{and where}
    \left\{ \begin{matrix}
    \dot{\overline{\delta}} \\ \dot{\overline{\varepsilon}}_1 \\
    \dot{\overline{\varepsilon}}_2 \\ \dot{\overline{\gamma}}_1 \\
    \dot{\overline{\gamma}}_2 \\ \dot{\overline{\gamma}}_3
    \end{matrix} \right\} & = \begin{bmatrix}
    1/3 & 1/3 & 1/3 & 0 & 0 & 0 \\
    1/3 & -1/3 & 0 & 0 & 0 & 0 \\
    0 & 1/3 & -1/3 & 0 & 0 & 0 \\
    0 & -\overline{\alpha} & \overline{\alpha} & 1 & 0 & 0 \\
    -\overline{\beta} & 0 & \overline{\beta} & \overline{\gamma} & 1 & 0 \\
    -\overline{\gamma} & \overline{\gamma} & 0 & 0 & 0 & 1
    \end{bmatrix} \left\{ \begin{matrix}
    \dot{\mathcal{V}}_{1i} \mathcal{V}^{-1}_{i1} \\ 
    \dot{\mathcal{V}}_{2i} \mathcal{V}^{-1}_{i2} \\ 
    \dot{\mathcal{V}}_{3i} \mathcal{V}^{-1}_{i3} \\
    \dot{\mathcal{V}}_{2i} \mathcal{V}^{-1}_{i3} \\ 
    \dot{\mathcal{V}}_{1i} \mathcal{V}^{-1}_{i3} \\ 
    \dot{\mathcal{V}}_{1i} \mathcal{V}^{-1}_{i2} 
    \end{matrix} \right\} \\
    \intertext{with}
    \dot{\overline{\varepsilon}}_3 & = -\dot{\overline{\varepsilon}}_1 - \dot{\overline{\varepsilon}}_2 .
    \end{align}
\end{subequations}
These strain rates can be integrated to get the Eulerian thermo\-dynamic strains $\overline{\delta}$, $\overline{\varepsilon}_1$, $\overline{\varepsilon}_2$, $\overline{\varepsilon}_3$, $\overline{\gamma}_1$, $\overline{\gamma}_2$ and $\overline{\gamma}_3$ by using initial conditions of  $\overline{\delta} |_0 = \overline{\varepsilon}_1 |_0 = \overline{\varepsilon}_2 |_0 = \overline{\varepsilon}_3 |_0 = \overline{\gamma}_1 |_0 = \overline{\gamma}_2 |_0 = \overline{\gamma}_3 |_0 = 0$ provided that the initial elongation ratios have been specified as $\overline{a}_0$, $\overline{b}_0$ and $\overline{c}_0$ and that the initial magnitudes of shear have been specified as $\overline{\alpha}_0$, $\overline{\beta}_0$ and $\overline{\gamma}_0$.

At this juncture, constitutive equations between the Eulerian thermo\-dynamic conjugate pairs $( \overline{\pi} , \overline{\delta} )$, $( \overline{\sigma}_1 , \overline{\varepsilon}_1 )$, $( \overline{\sigma}_2 , \overline{\varepsilon}_2 )$, $( \overline{\tau}_1 , \overline{\gamma}_1 )$, $( \overline{\tau}_2 , \overline{\gamma}_2 )$ and $( \overline{\tau}_3 , \overline{\gamma}_3 )$ are to be introduced (again, a topic for future work) to solve for the Eulerian thermo\-dynamic stresses $\overline{\pi}$, $\overline{\sigma}_1$, $\overline{\sigma}_2$, $\overline{\tau}_1$, $\overline{\tau}_2$ and $\overline{\tau}_3$.  After the thermo\-dynamic stresses have been updated they can be mapped back into the components of Kirchhoff stress $\tau_{ij}$ via
\begin{equation}
\left\{ \begin{matrix}
\tau_{11} \\ \tau_{22} \\ \tau_{33} \\
\tau_{23} = \tau_{32} \\ \tau_{13} = \tau_{31} \\ \tau_{12} = \tau_{21}
\end{matrix} \right\} = \begin{bmatrix}
1/3 & 2/3 & 1/3 & -\overline{\alpha} & 0 & -2\overline{\gamma} \\
1/3 & -1/3 & 1/3 & -\overline{\alpha} & 0 & \overline{\gamma} \\
1/3 & -1/3 & -2/3 & 2\overline{\alpha} & 0 & \overline{\gamma} \\
0 & 0 & 0 & 1 & \overline{\gamma} & 0 \\
0 & 0 & 0 & 0 & 1 & 0 \\
0 & 0 & 0 & 0 & 0 & 1
\end{bmatrix} \left\{ \begin{matrix}
\overline{\pi} \\ \overline{\sigma}_1 \\ \overline{\sigma}_2 \\
\overline{\tau}_1 \\ \overline{\tau}_2 \\ \overline{\tau}_3
\end{matrix} \right\}
\end{equation}
from which any commonly used stress tensor can be easily gotten.

Although $\overline{\sigma}_3$ and $\dot{\overline{\varepsilon}}_3$ are not needed from a constitutive perspective, they are required to calculate stress power.  Also,
to correctly compute stress power, Eqns.~(\ref{EulerianStressPower1} or \ref{EulerianStressWorking} \& \ref{EulerianStressPower2}) must both contribute, the former because of straining and the latter because of co\-ordinate spin.  A numerical strategy based upon quaternion theory to acquire spin tensors from rotation tensors by using finite difference schemes can be found in Ref.~\cite{Freedetal19}.

\section{Conclusions}

Lagrangian and Eulerian triangular decompositions of deformation have been analyzed and compared.  Physically observable stretch\slash strain components comprising the triangular Laplace stretch of each decomposition have been derived and then highlighted in several example problems involving homogeneous deformations.  Consideration of stress power, i.e., rate of working done by each stretch rate, has enabled derivation of work conjugate stress-stretch tensors as-well-as thermo\-dynamically conjugate scalar pairs of stress-strain attributes with physical meaning.  Significantly, the Eulerian formulation containing an Eulerian, lower-triangular, stretch tensor has not been developed elsewhere in the mechanics literature, to the authors' knowledge.  The current results provide a theoretical foundation for construction of constitutive models to be undertaken in future work.

\medskip\noindent\textbf{Acknowledgement}\medskip

This research was inspired by a conversation that ADF had with Prof.\ Michael Sacks at the 2019 Annual Meeting of the Society for Engineering Science held at Washington University in St.\ Louis, where he encouraged the author to develop of an Eulerian constitutive framework suitable for bio\-mechanics.

SZ was funded by the U.S.\ Army Research Laboratory, Aberdeen, MD.


\end{document}